\documentclass[twocolumn,preprintnumbers,amsmath,amssymb,a4paper,aps,pra,10pt]{revtex4-1}
\usepackage{amssymb}
\usepackage{amsfonts}
\usepackage{amsmath}
\usepackage{amsthm}
\usepackage{graphicx}
\usepackage{braket}
\usepackage{bm}
\usepackage{epstopdf}

\begin{document}

\author{X.-W. Luo$^{1,2}$, J. J. Hope$^{3}$, %G. J. Milburn$^{1}$,
 B. Hillman$^{3}$, and T. M. Stace$^{1}$}\email[]{stace@physics.uq.edu.au}
\affiliation{$^1$ARC Centre for Engineered Quantum Systems, University of Queensland, St Lucia, QLD 4072,Australia}
\affiliation{$^2$Key Lab of Quantum Information, CAS,
University of Science and Technology of China,
Hefei, Anhui, 230026, P.R. China}
\affiliation{$^3$Department of Quantum Science,
Research School of Physics and Engineering,
The Australian National University, Canberra, ACT0200, Australia}
\title{\textbf{Diffusion Effects in Gradient Echo Memory}}
% insert the table of contents
%\tableofcontents
\begin{abstract}
We study the effects of diffusion on a $\Lambda$-gradient echo memory, which is a coherent optical quantum memory using thermal gases.  The efficiency of this memory is high for short storage time, but decreases exponentially due to decoherence as the storage time is increased. We study the effects of both longitudinal and transverse diffusion in this memory system, and give both analytical and numerical results that are in good agreement. Our results show that diffusion has a significant effect on the efficiency. Further, we suggest ways to reduce these effects to improve storage efficiency.
\end{abstract}

\maketitle

\section{Introduction}
Quantum memory is an important tool in many quantum information protocols, including quantum repeaters for long-distance quantum communication \cite{duan2001long}, and identity quantum gates in quantum computation \cite{knill2001scheme}. Numerous optical quantum memories have been developed, including electromagnetically-induced transparency (EIT) based quantum memory \cite{fleischhauer2005electromagnetically, fleischhauer2000dark}, far-detuned Raman process memory \cite{nunn2006modematching, nunn2007mapping}, and photon-echo quantum memories: controlled reversible inhomogeneous broadening (CRIB) memory \cite{kraus2006quantum, sangouard2007analysis}, atomic frequency combs (AFC) memory \cite{afzelius2009multimode}, and gradient echo memory (GEM) \cite{hetet2006gradient, hetet2008electro, hetet2008quantum}. A review of these schemes can be found in \cite{lvovsky2009optical}. Of these schemes, the most impressive efficiency so far attained experimentally is 87\% by $\Lambda$-GEM scheme \cite{hosseini2011high} using warm rubidium vapor.  In this paper, we will examine the effects of atomic diffusion on the $\Lambda$-GEM system, which may limit this efficiency for larger storage times.

\begin{figure}
\includegraphics[angle=0, width=0.25\textwidth]{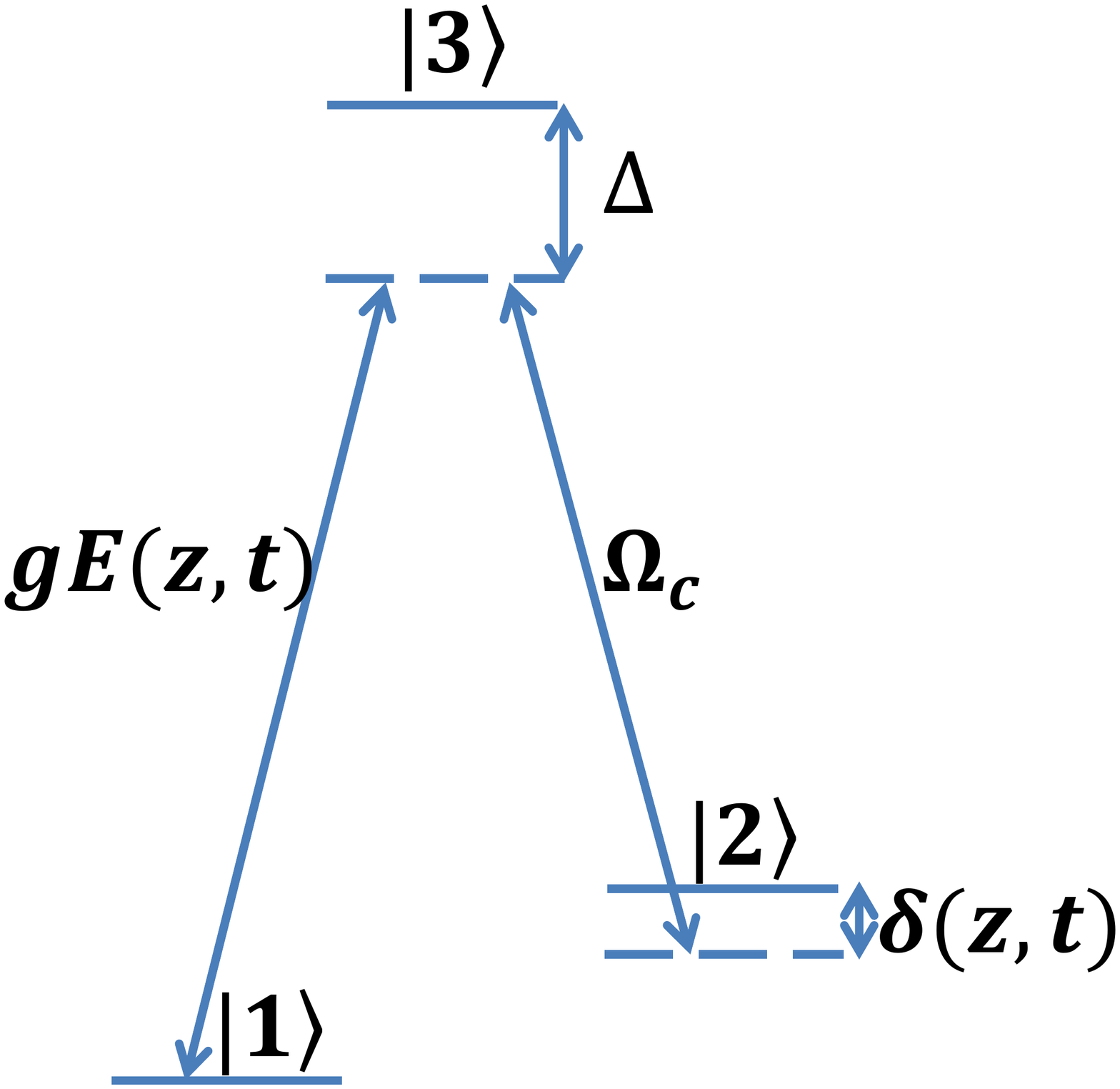}
\caption{\footnotesize{Level structure of $\Lambda$-type 3-level atom.}}
\label{fig:3-lvl}
\end{figure}

$\Lambda$-GEM is a memory using a 3-level, $\Lambda$-type atom (Fig. \ref{fig:3-lvl}). The input optical pulse couples the two metastable lower states through a control field. The excited state is coupled in the far detuned region, so the 3-level atoms can be treated as effective 2-level atoms.  These effective two-level atoms have linearly increasing atomic Zeeman shifts along the length of the storage medium. The pulse is first absorbed, then by simply reversing the sign of the magnetic field, the pulse is retrieved in the forward direction.  The incident signal field is converted to a collective atomic excitation known as a spin wave, which is distributed as a function of position. The Brownian motion of the gaseous atoms will cause diffusion, which will disturb spatial coherence of the atomic spin-wave, leading to decoherence. For a plane wave, only axial diffusion is important, but transverse diffusion becomes significant when a realistic beam profile is included. There has been recent interest in the effects of diffusion in the EIT \cite{firstenberg2008theory, firstenberg2009elimination, firstenberg2010self}. In this work, we study the effects of diffusion in $\Lambda$-GEM system, giving analytical and numerical results. We also suggest ways to reduce these effects to improve storage efficiency.

\section{$\Lambda$ Gradient Echo Memory}
We consider a medium consisting of $\Lambda$-type 3-level atoms with two metastable lower states as shown in Fig.~\ref{fig:3-lvl}.
The ground state $\vert 1\rangle$ and the excited state $\vert 3 \rangle$ are coupled by a weak optical field, the positive frequency component of the electric field is described by the slowly varying operator
\begin{equation}
\hat{E}(\textbf{r},t)=\sum_{\textbf{k}}\sqrt{\frac{1}{V}} a_{\textbf{k}}(t)e^{i\textbf{k}\cdot \textbf{r}}e^{-ik_0z}e^{i\omega_0 t}
\end{equation}
with detuning $\Delta$,
where $V$ is the quantization volume, $\omega_0$ is the carrier frequency of the quantum field and $k_0=\omega_0/c$. The excited state $\vert 3 \rangle$ is also coupled to the metastable state $\vert 2 \rangle$ via a coherent control field with Rabi frequency $\Omega_c$ and a two photon detuning $\delta$. This two photon detuning is spatially varied $\delta(z,t)=\eta(t) z$, with time dependent gradient $\eta(t)$. Then the interaction Hamiltonian in the rotating frame with respect to the field frequencies is
\begin{equation}
\begin{split}
\hat{H}=&\sum_{n}[\hslash \Delta \sigma_{33}^{(n)} + \hslash \delta(z_n,t) \sigma_{22}^{(n)} \\
+ &\hslash \sum_{\textbf{k}}(\hslash g_{\textbf{k}} a_{\textbf{k}} e^{i\textbf{k}\cdot \textbf{r}_n} \sigma_{31}^{(n)} + \hslash \Omega_c(\textbf{r}_n) \sigma_{32}^{(n)} + h.c)],
\end{split}
\end{equation}
where $g_{\textbf{k}}=\wp \sqrt{\frac{\omega_{\textbf{k}}}{2\hslash \epsilon_0 V}}$ is the atom-field coupling constant with $\wp$ being the dipole moment of the 1-3 transition, and $\sigma_{\mu \nu}^{(n)}=\vert \mu \rangle_n \langle \nu \vert$ is an  operator acting on the $n$-th atom at $\textbf{r}_n=(x_n,y_n,z_n)$.
We assume that initially all atoms are in their ground state $\vert 1\rangle$. We transform to  collective operators, which  are averages over atomic operators over a small volume centered at $\textbf{r}$ containing $N_{\textbf{r}} \gg 1$ particles,
\begin{equation}
\sigma_{\mu \nu}(\textbf{r},t)=\frac{1}{N_{\textbf{r}}}\sum_{j=1}^{N_{\textbf{r}}}\sigma^{(j)}_{\mu \nu}(t)
\end{equation}
From the Heisenberg-Langevin equations in the weak probe region ($\sigma_{11}\simeq 1, \sigma_{22}\simeq\sigma_{33}\simeq 0$), we get the Maxwell-Bloch equations \cite{walls2008quantum},
\begin{equation}\label{M-B-1}
\begin{split}
&\dot{\sigma}_{13}^{(n)}=-(\gamma_{13}+i\Delta)\sigma_{13}^{(n)}+ige^{ik_0z_n}E(\textbf{r}_n,t)+i
\Omega_c
e^{ik_cz_n}\sigma_{12}^{(n)},\\
&\dot{\sigma}_{12}^{(n)}=-(\gamma_{12}+i\delta(z_n,t))\sigma_{12}^{(n)}+
i\Omega_c e^{-ik_cz_n}\sigma_{13}^{(n)},\\
&\left( \frac{\partial}{\partial t}+c\frac{\partial}{\partial z}-ic\frac{\nabla^2_x+\nabla^2_y}{2k_0}\right)E(\textbf{r},t)
=igNe^{-ik_0z}\sigma_{13}(\textbf{r},t),
\end{split}
\end{equation}
%or equally
%\begin{equation}
%\begin{split}
%\dot{\sigma}_{13}(z,t) =& -(\gamma_{13}+i\Delta)\sigma_{13}(z,t)\\
%&+ige^{ik_0z}E(z,t)+i\Omega_c(t)e^{ik_cz}\sigma_{12}(z,t),\\
%\dot{\sigma}_{12}(z,t)=&-(\gamma_{12}+i\delta(z,t))\sigma_{12}(z,t)\\
%&+i\Omega_c(t)e^{-ik_cz}\sigma_{13}(z,t),\\
%\left( \frac{\partial}{\partial t}+c\frac{\partial}{\partial z}\right)&E(z,t)
%=igNe^{-ik_0z}\sigma_{13}(z,t).
%\end{split}
%\end{equation}
where $\gamma_{\nu \mu}$ are the decay rates, $g=\wp \sqrt{\frac{\omega_0}{2\hslash \epsilon_0}}$, and $N$ is the atomic density. We have assumed that $g_{\textbf{k}}\simeq\wp \sqrt{\frac{\omega_0}{2\hslash \epsilon_0 V}}$ and $\Omega_c(\textbf{r})=\Omega_c e^{ik_cz}$. We also omit the Langevin noise operators since here we are more interested in the decoherence caused by diffusion. This is equivalent to making a semiclassical approximation for the electric field and the atomic coherences.

In Eq.~(\ref{M-B-1}), $ic\frac{\nabla^2_x+\nabla^2_y}{2k_0}$ is the diffraction term, and generally, the diffraction effects can be neglected \cite{stace2010theory}. Notice that we are here considering the regime $t_p\gg L/c$, where $2t_p$ is the temporal width of the signal, and $2L$ is the length of the medium. This allows us to neglect temporal retardation effects, i.e., we can neglect the temporal derivative in the third equation of Eq.~(\ref{M-B-1}).
Also, since the atoms are far detuned (\ $\Delta\gg \gamma_{13}, \Omega_c$), we adiabatically eliminate the fast oscillations and set $\dot{\sigma}_{13}^{(n)}=0$. Then we have $\sigma_{13}= \left(ge^{ik_0z}E+\Omega_ce^{ik_cz}\sigma_{12} \right)/\Delta$, and we get the reduced Maxwell-Bloch equations,
\begin{equation}\label{M-B-2}
\begin{split}
\dot{\sigma}_{12}^{(n)}=&-(i\delta(z_n,t)-i\frac{\Omega_c^2}{\Delta})\sigma_{12}^{(n)}\\
&+i\frac{g\Omega_c}{\Delta}e^{i(k_0-k_c)z_n}E(\textbf{r}_n,t),\\
\frac{\partial}{\partial z}E(\textbf{r},t)=&i\frac{gN\Omega_c}{c\Delta}  e^{-i(k_0-k_c)z}\sigma_{12}(\textbf{r},t)\\
&+i\frac{g^2N}{c \Delta}E(\textbf{r},t).
\end{split}
\end{equation}
Here we neglect decay, i.e.\ $\gamma_{12}\rightarrow0$, since we consider the storage time much less than $1/\gamma_{12}$.

\section{Diffusion}
We now consider the effects of diffusion on the atomic state. In order to isolate the motional effects of  diffusion from collisional dephasing, we assume that the collisions between atoms do not change the state of the atom.  Then we derive the diffusion equation for the atomic density matrix $\rho$. Space is divided into volume elements with length $\Delta r$ and center $r$. We associate a density matrix $\rho(r,t)$ with atoms in this volume element, given by
\[\rho(r,t)=\frac{1}{N_r}\sum_{j=1}^{N_r}\rho^{(j)}(t),\]
where $N_r$ is the atom number in volume centered at $r$. The total density matrix for the entire system is assumed to be the tensor product of these local density matrices.

Diffusion causes an exchange of atoms between adjacent volumes. During a short time $\Delta t$, a fraction $\epsilon$ of the atoms in slice $r$ migrate into slice $r\pm\Delta r$. There is also atomic flux back into slice $r$ from $r\pm\Delta r$. We assume that the total number density of the atoms is uniform, so the state at $r$ and $t+\Delta t$ is described by the new density matrix which is the average of the density matrix of atoms remaining in the volume and those that have migrated in to it. The diffusive component of the evolution is therefore
\begin{equation}
\begin{split}
\rho(r,t+\Delta t)=&(1-2\epsilon)\rho(r,t)\\
&+\epsilon(\rho(r+\Delta r,t)+\rho(r-\Delta r,t))\\
\Rightarrow \partial_t\rho(r,t)=&D\nabla^2\rho(r,t)
\end{split}
\end{equation}
where $D=\epsilon \Delta r^2/\Delta t$ is the diffusion coefficient.
With the same consideration, we get the diffusive component evolution for the atomic correlation functions
\begin{equation}
\dot{\sigma}_{\mu \nu}(r,t)=D\nabla^2\sigma_{\mu \nu}(r,t)
\end{equation}
Now we introduce the interaction with optical fields.  Since diffusion is caused by Brownian motion, this will lead to Doppler shifts in the various detunings.  We now consider the interaction between the optical field and a single atom, and quantify the effects of these Doppler shifts. The atom moves at some random velocity, and there will be a Doppler shift for both the signal and control fields. So the detunings in Eq.~(\ref{M-B-1}) become $\Delta=\Delta_0+\Delta_{Dopp}$,  and $\delta=\delta_0+\delta_{Dopp}$, with $\Delta_0, \delta_0$ the detunings for stationary atoms and $\Delta_{Dopp}, \delta_{Dopp}$ the Doppler shifts. Typically, the one photon Doppler shift $\Delta_{Dopp}\ll\Delta_0$, state $\vert 3\rangle$ is still far detuned. So the adiabatic elimination is still valid in the presence of the  Brownian motion induced Doppler shift,  and  we can still reduce the 3-level atom to an effective 2-level atom. The Maxwell-Bloch equation will still reduce to Eq.~(\ref{M-B-2}), but with one photon detuning $\Delta=\Delta_0+\Delta_{Dopp}$ and two photon detuning $\delta=\delta_0+\delta_{Dopp}$. So, for the reduced two level atomic system, the diffusive Maxwell-Bloch equation for the collective correlation $\sigma_{12}(z,t)$ averaged over atoms in each volume is
\begin{equation}\label{M-B-tr}
\begin{split}
\dot{\sigma}_{12}(\textbf{r},t)=&i\frac{g\Omega_c}{\Delta}e^{i(k_0-k_c)z}E(\textbf{r},t)\\
&-i\delta(z,t)\sigma_{12}(\textbf{r},t)+D\nabla^2\sigma_{12}(\textbf{r},t),\\
\frac{\partial}{\partial z}E(\textbf{r},t)=&
i\frac{gN\Omega_c}{c\Delta}  e^{-i(k_0-k_c)z}\sigma_{12}(\textbf{r},t)\\
&+i\frac{g^2N}{c\Delta}E(\textbf{r},t).
\end{split}
\end{equation}
We have absorbed the Stark shift $\frac{\Omega_c^2}{\Delta}$ into the two-photon detuning. Here our diffusive
Maxwell-Bloch equation is consistent with the result in the EIT system \cite{firstenberg2008theory, firstenberg2010self}.
%So we don't need to consider the diffusion of $\sigma_{13}(z,t)$. Though spatially $\sigma_{13}(z,t)$ oscillates very fast due to the phase factor $e^{ik_0z}$ in the coupling with signal field, the control field will couple $\sigma_{13}$ to $\sigma_{12}$ in a phase matched manner, and will cancel the phase $e^{ik_0z}$ by another phase factor $e^{-ik_cz}$, the phase will always match even when the atom is under a random moving. That is why we can make the adiabatic elimination and neglect the diffusion of $\sigma_{13}(z,t)$.

Notice that the signal and control fields are co-propagating, so the Doppler broadening width for $\delta$ is typically $1$kHz, which is much smaller than the frequency width of the signal field ($\sim$ 1MHz), so we neglect this two-photon Doppler broadening $\delta_{Dopp}$ and replace $\delta(z,t)$ by $\delta_0(z,t)=\eta(t) z, z\in[-L,L]$.

For the one photon detuning $\Delta=\Delta_0+\Delta_{Dopp}$, after we make the adiabatic elimination, it will appear in the denominator (see Eq.~(\ref{M-B-tr})), so
\[\frac{1}{\Delta}\simeq\frac{1}{\Delta_0}\left(1-\frac{\Delta_{Dopp}}{\Delta_0}+\left(\frac{\Delta_{Dopp}}{\Delta_0}\right)^2\right).\] The term linear in $\Delta_{Dopp}$  will vanish when we average over many atoms in a volume centred at $r$, so we can replace $\Delta$ by $\Delta_0$ in our Maxwell-Bloch equation, with second order accuracy [typically $(\Delta_{Dopp}/\Delta_0)^2\sim10^{-3}$].

%In this paper we will choose the parameters such that the optical depth is sufficient
%large, which will lead to 100\% efficiency if there is no decoherence.
%Thus we  can focus on the diffusion effects, and don't need to consider the influence of
%optical depth on efficiency.

\section{Analytic Calculation and Numerical Simulation}

To quantify the effects of diffusion, we solve  for the atomic dynamics.  There are three distinct phases during the storage: write-in  $-t_0<t<0$, during which the signal is  absorbed by the memory; hold $0<t<t_H$, during which the information is stored in the memory and the gradient is turned off; read-out $t_H<t<t_H+t_0$, during which the signal is emitted by turning on the flipped gradient.

We quantify the effects of diffusion by the read-out efficiency $\varepsilon$ defined to be
\begin{equation}\label{eq:eff1}
\varepsilon=\frac{\int_{t_H}^{t_H+t_0}|f_{out}(t)|^2 dt}{\int_{-t_0}^{0}|f_{in}(t)|^2 dt}
\end{equation}
where $f_{out}(t)=E(z=L,t>t_H)$ is the output field and $f_{in}(t)=E(z=-L,t<0)$ is the input field. We solve for $f_{out}(t)$ both numerically and analytically, and consider the effects of diffusion in axial (longitude) and radial (transverse) directions separately.

\subsection{Longitudinal diffusion}
For a uniform plane wave, transverse diffusion is irrelevant. We replace $\textbf{r}$ by $z$ in Eq.~(\ref{M-B-tr}) and consider the longitude diffusion in a 1-dimensional model. Now the Maxwell-Bloch equation is
\begin{equation}\label{M-B-Eq}
\begin{split}
\dot{\sigma}_{12}(z,t)=&i\frac{g\Omega_c}{\Delta}e^{i(k_0-k_c)z}E(z,t)\\
&-i\delta(z,t)\sigma_{12}(z,t)+D\nabla_z^2\sigma_{12}(z,t),\\
\frac{\partial}{\partial z}E(z,t)=&
i\frac{gN\Omega_c}{c\Delta}  e^{-i(k_0-k_c)z}\sigma_{12}(z,t)\\
&+i\frac{g^2N}{c\Delta}E(z,t).
\end{split}
\end{equation}
We now investigate the longitude diffusion effects during the write-in process, the hold time and the read-out processes separately.

To compute $f_{out}(t)$, we evolve Eq.~(\ref{M-B-Eq}) using $\eta$ as in Fig. \ref{fig:k-t} (a). Following the method given in \cite{hetet2006gradient}, we first propagate
$E(z,t)$ and $\sigma_{12}(z,t)$ forward with boundary condition $E(z=-L,t<0)=f_{in}(t)$ to find their values at time $t_H$. Then we propagate $E$ and $\sigma_{12}$ backward to time $t_H$, with final condition $E(z=L,t>t_H)=f_{out}(t)$, and solve for $f_{out}(t)$ by matching the two solutions at time $t_H$.

%To solve Eq.~(\ref{M-B-Eq}), we first propagate $E(z,t)$ and $\sigma_{12}(z,t)$ forward
%with boundary condition $\sigma_{12}(z,t\rightarrow-\infty)=0$ and $E(z=-L,t<0)=f_{in}(t)$
%to find their values in the region ($t\in [-t_0,0]$, $z\in [-L,L]$). Then propagate
%$\sigma_{12}(z,0)$ to $\sigma_{12}(z,t_H)$. To find the values in the region
%($t\in[t_H,t_H+t_0]$, $z\in [-L,L]$), we propagate $\sigma_{12}$ and $E$ backward with final
%condition $\sigma_{12}(z,t\rightarrow+\infty)=0$ and $E(z=L,t>t_H)=f_{out}(t)$. By
%matching the value at $t_H$, we can get the expression for the output pulse $f_{out}(t)$.

\begin{figure}
\includegraphics[angle=0, width=0.5\textwidth]{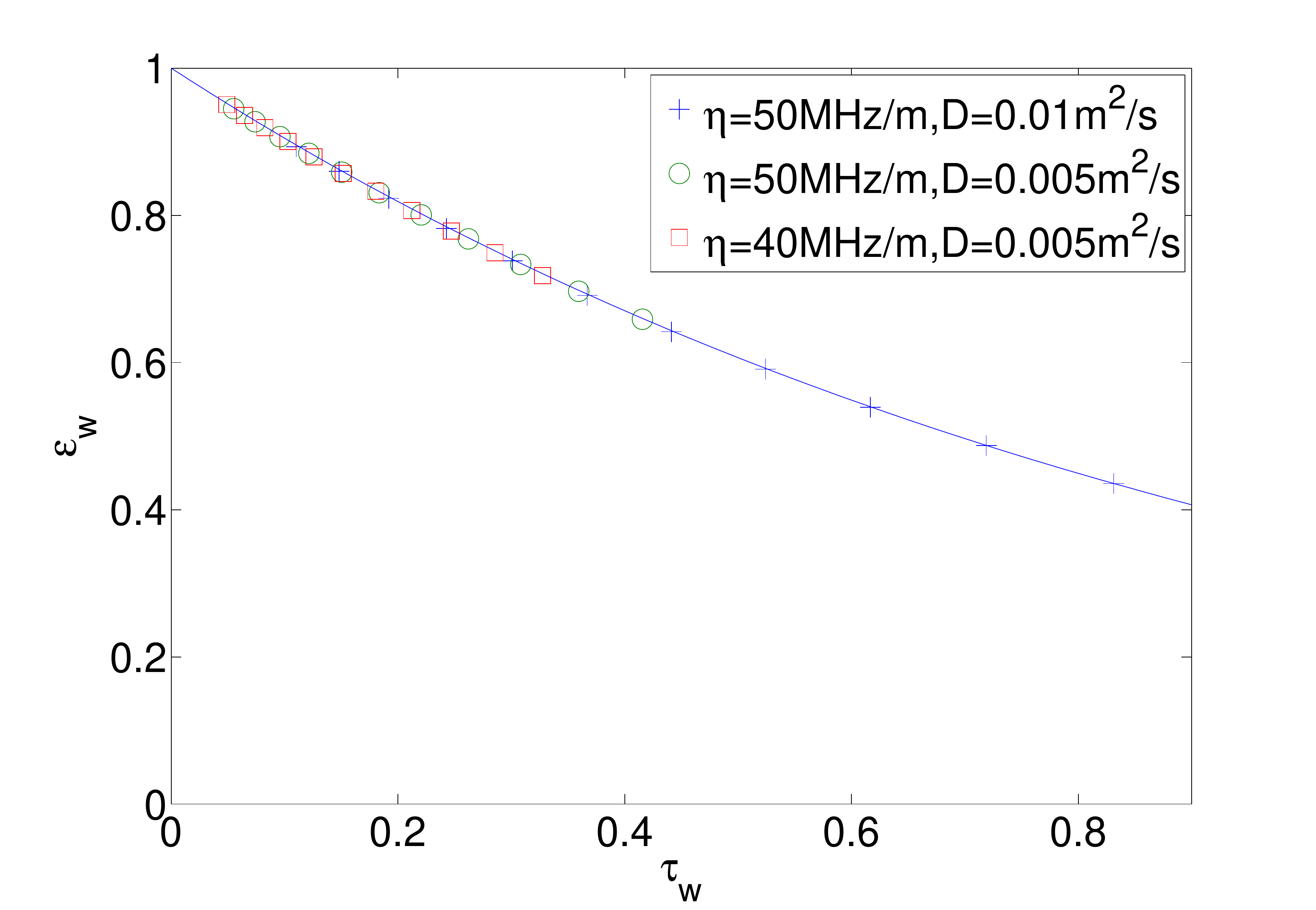}
\caption{\footnotesize{The efficiency decay with respect to the dimensionless parameter $\tau_{W}$ for longitude diffusion during the write-in time, the points are numerical results, and the curve is Eq.~(\ref{eq:eff-w}).}}
\label{fig:1D_input}
\end{figure}

\emph{Write:} Consider the diffusion effects during the write-in process, we find that (see Appendix A)
\begin{equation}
f_{out}(t_H+t)=e^{\frac{-D}{3\eta}(k_i^3-(k_i-\eta t)^3)}f_{in}(-t)\bar{G}
\end{equation}
where $k_i=\frac{g^2N}{c\Delta}+k_0-k_c-\frac{\beta}{L}$ is the initial spatial frequency of $\sigma_{12}(z,t)$, and \[\bar{G}=\vert\eta L\left(t+\frac{\beta}{\eta L}\right)\vert ^{-i2\beta}e^{i\frac{2Lg^2N}{c\Delta}}e^{-i\frac{g_{eff}^2N}{c\eta \left(t+\frac{\beta}{\eta L}\right)}t_H}\Gamma(i\beta)/\Gamma(-i\beta)\] is a phase factor, with $g_{eff}=g\Omega_c/\Delta$ and $\beta=\frac{g_{eff}^2N}{\eta c}$.
For a pulse with Gaussian temporal profile $f_{in}=Ae^{-(t+t_{in})^2/t_p^2}$, we find
\begin{equation}
\varepsilon_{W}=\frac{\int_{-t_0}^0 dte^{-2(t+t_{in})^2/t_p^2} e^{\frac{-2D}{3\eta}(k_i^3-(k_i+\eta t)^3)}}{\int_{-t_0}^0 dte^{-2(t+t_{in})^2/t_p^2}}.
\end{equation}
Typically, $D\eta^2t_p^3$ is very small,
%and to the second order of $D\eta^2t_p^3$,
then the efficiency is
\begin{equation}\label{eq:eff-w-1}
\varepsilon_{W}=\sqrt{\alpha_{W}} e^{-\tau_{W}}+O(D^2\eta^4t_p^6)
\end{equation}
where $\alpha_{W}=\frac{1}{1-D\eta^2t_p^2(k_i/\eta-t_{in})}$ and $\tau_{W}=\frac{2D\eta^2}{3}\left[\left(\frac{k_i}{\eta}\right)^3-\left(\frac{k_i}{\eta}-t_{in}\right)^3\right]$ are dimensionless parameters.
For typical experimental parameters, $\alpha_{W}\simeq 1$, then
\begin{equation}\label{eq:eff-w}
\varepsilon_{W}\simeq e^{-\tau_{W}}
\end{equation}
We also numerically solve Eq.~(\ref{M-B-Eq}) with diffusion during the write-in process, using XMDS \cite{XMDS}. We calculate the efficiency for different values of the diffusion rate $D$, input time $t_{in}$ etc. The results are shown in Fig. \ref{fig:1D_input}, (points are numerical results, and the curve is Eq.~(\ref{eq:eff-w})). We plot the efficiency $\varepsilon_{W}$ with respect to the rescaled dimensionless parameter $\tau_{W}$, so all the points with different parameters collapse on a single curve.

\begin{figure}
\includegraphics[angle=0, width=0.5\textwidth]{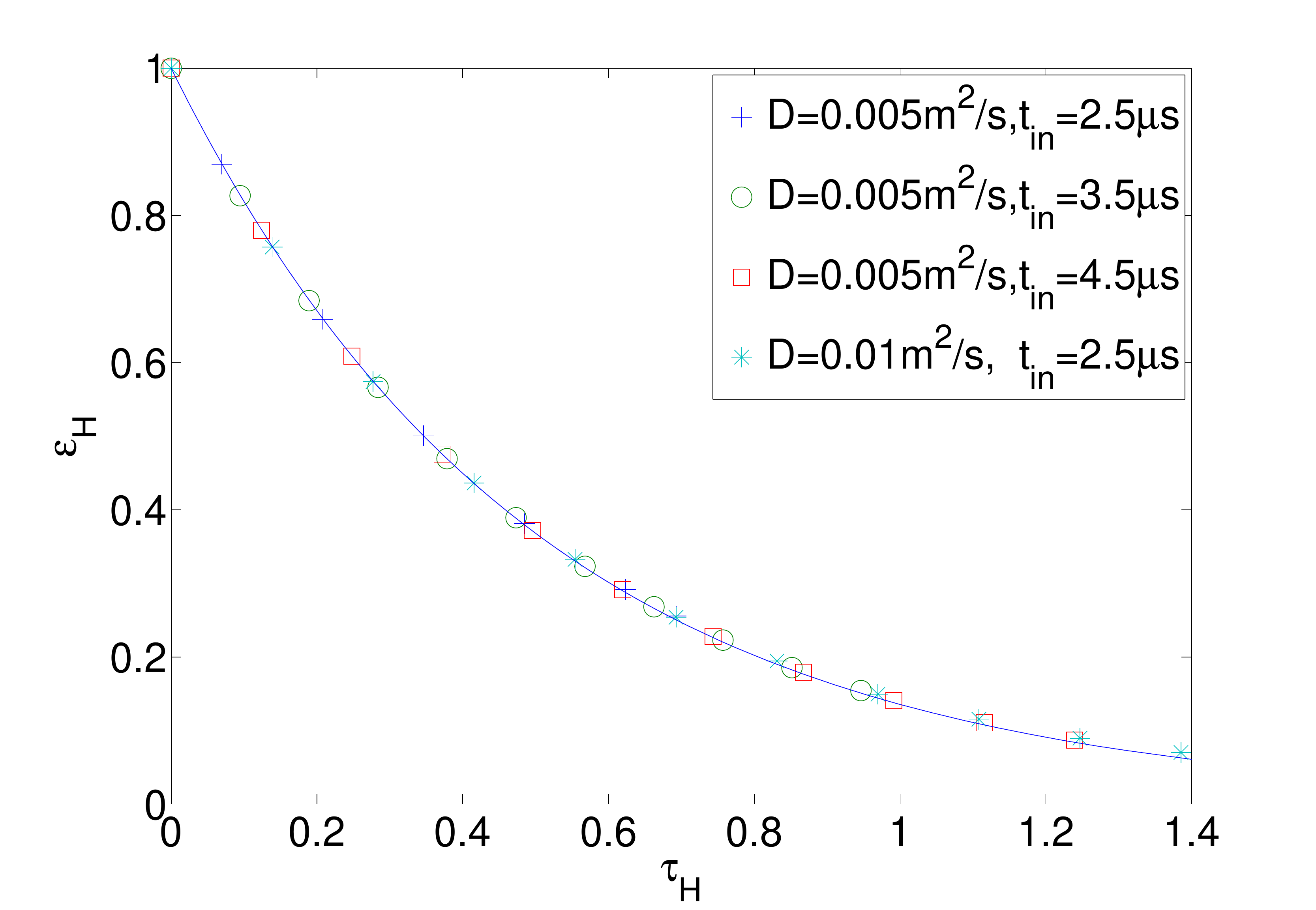}
\caption{\footnotesize{The efficiency decay with respect to the dimensionless parameter $\tau_H$ for longitude diffusion during hold time, the points are numerical results, and the curve is Eq.~(\ref{eq:eff-H}).}}
\label{fig:1D_hold}
\end{figure}

\emph{Hold:} During the storage time $[0,t_H]$,  we find (see Appendix A)
\begin{equation}
f_{out}(t_H+t)=e^{-Dt_H(k_i-\eta t)^2 }f_{in}(-t)\bar{G}
\end{equation}
For the above Gaussian shape input, the efficiency is given by
\begin{equation}\label{eq:eff-hold}
\varepsilon_H=\sqrt{\alpha_H}e^{-2\alpha_H \tau_H},
\end{equation}
where $\alpha_H=\frac{1}{t_p^2}/\left(\frac{1}{t_p^2}+Dt_H\eta^2\right)$ and $\tau_H=Dt_Hk_H^2$ are dimensionless parameters, with $k_H=k_i-\eta t_{in}$.
For typical experimental parameters, $\alpha_H\simeq 1$ and we have
\begin{equation}\label{eq:eff-H}
\varepsilon_H\simeq e^{-2 \tau_H }.
\end{equation}
We also numerically solve  Eq.~(\ref{M-B-Eq})  with diffusion during hold time, using XMDS.
We calculate the efficiency for different values of the diffusion rate $D$, storage time $t_H$ etc. The results are shown in Fig. \ref{fig:1D_hold} (points are numerical results, and curve is Eq.~(\ref{eq:eff-H})).  We plot the efficiency $\varepsilon_H$ with respect to the rescaled dimensionless parameter $\tau_H$, so all the points with different parameters collapse on a single curve.

\emph{Read:} the diffusion effects during the read-out process are the same as the diffusion effects of the write-in process (see the appendix A), so we simply have\[\varepsilon_{R}=\varepsilon_{W}.\]
\subsection{Transverse diffusion}
We now quantify the effects of diffusion for a beam with realistic transverse Gaussian profile.
%\begin{equation}
%\left( \frac{\partial}{\partial t}+c\frac{\partial}{\partial z}-ic\frac{\nabla^2_x+\nabla^2_y}{2k_0}\right)E(\textbf{r},t)
%=igNe^{-ik_0z}\sigma_{13}(\textbf{r},t)
%\end{equation}
%where $ic\frac{\nabla^2_x+\nabla^2_y}{2k_0}$ is the diffraction term. Generally, the diffraction effects can be neglected. Then, according to the same arguments as before, and assume the control field is homogeneous, we  have
%\begin{equation}\label{M-B-tr}
%\begin{split}
%\dot{\sigma}_{12}(\textbf{r},t)=&-(i\eta z)\sigma_{12}(\textbf{r},t)+i\frac{g\Omega_c}%{\Delta}e^{i(k_0-k_c)z}E(\textbf{r},t)\\
%&+D\nabla^2\sigma_{12}(\textbf{r},t),\\
%\frac{\partial}{\partial z}E(\textbf{r},t)
%=&i\frac{gN\Omega_c}{c\Delta}  e^{-i(k_0-k_c)z}\sigma_{12}(\textbf{r},t)\\
%&+i\frac{g^2N}{c\Delta}E(\textbf{r},t).
%\end{split}
%\end{equation}
The efficiency for a 3-dimensional model is defined as
\begin{equation}\label{eq:3d-eff}
\varepsilon=\frac{\int|f_{out}(x,y,t_H+t)|^2 dxdydt}{\int|f_{in}(x,y,t)|^2 dxdydt}
%&=\frac{\int |f_{out}(k_x,k_y,t_H+t)|^2 dk_xdk_ydt}{\int |f_{in}(k_x,k_y,t)|^2 dk_xdk_ydt}
\end{equation}
Eq.~(\ref{M-B-tr}) can be solved in Fourier space $k_x,k_y$, and also notice that
\begin{equation}
\varepsilon=\frac{\int |f_{out}(k_x,k_y,t_H+t)|^2 dk_xdk_ydt}{\int |f_{in}(k_x,k_y,t)|^2 dk_xdk_ydt}
\end{equation}

Eq.~(\ref{M-B-tr}) can be reduced to a quasi-1D problem, and can be solved as before (see Appendix B).
For transverse diffusion, the output pulse will be
\begin{equation}\label{eq:3d-out}
f_{out}(k_x,k_y,t_H+t)=e^{-2\gamma_k t}e^{-\gamma_k t_H}f_{in}(k_x,k_y,-t)\bar{G}.
\end{equation}
where $\gamma_k=D(k_x^2+k_y^2)$.

If the input pulse has both Gaussian temporal and transverse profile,
\[f_{in}(x,y,t)=Ae^{-(x^2+y^2)/a^2}e^{-(t+t_{in})^2/t_p^2},\]
then $\gamma_k t_p\sim Dt_p/a^2$, which is typically small.
Thus the memory efficiency is
\begin{equation}\label{eq:eff-tr}
\varepsilon_{\perp}=\frac{1}{1+\tau_{\perp}}+O(\gamma_k^2t_p^2),
\end{equation}
where $\tau_{\perp}=4D(t_H+2t_{in})/a^2$ is a dimensionless parameter.

\begin{figure}
\includegraphics[angle=0, width=0.5\textwidth]{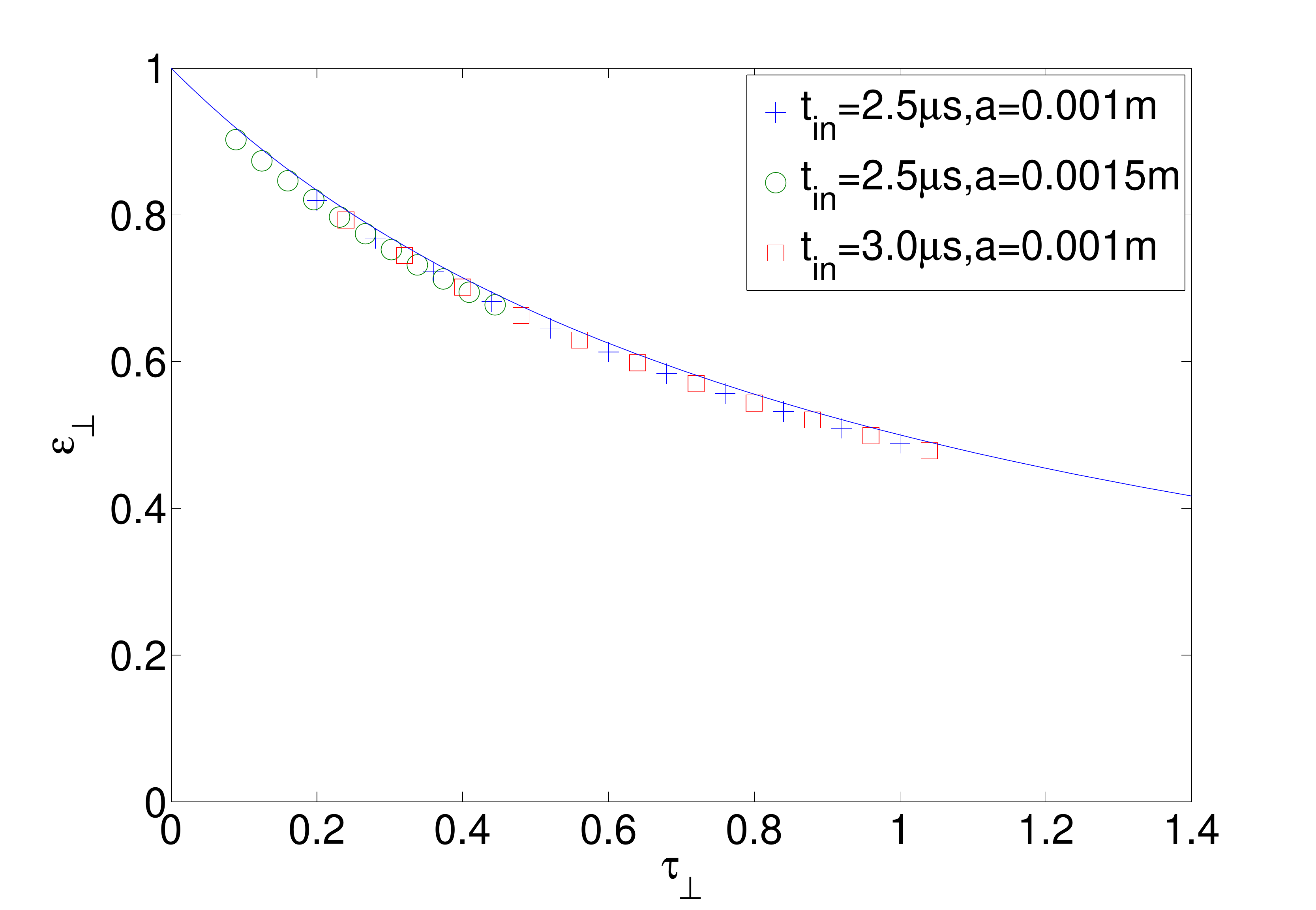}
\caption{\footnotesize{Efficiency decay with respect to $\tau_{\perp}$ for transverse diffusion, the points are numerical results, and the curve is Eq.~(\ref{eq:eff-tr}).}}
\label{fig:3D_trans}
\end{figure}

We also numerically solve  Eq.~(\ref{M-B-tr}) with $\nabla^2=\nabla_x^2+\nabla_y^2$.
We calculate the efficiency for different values of $a$, $t_H$ etc. The results are shown in Fig. \ref{fig:3D_trans} (points are numerical results, and curve is Eq.~(\ref{eq:eff-tr})). We plot the efficiency $\varepsilon_{\perp}$ with respect to the rescaled dimensionless parameter $\tau_{\perp}$, so all the points with different parameters collapse on a single curve.
\subsection{Total diffusion}
Experimentally, longitude and transverse diffusion coexist during the whole process.
Combining all the diffusive contributions mentioned above, we get the output field as (Appendix B)
\begin{equation}
\begin{split}
f_{out}(k_x,k_y,t_H+t)=&e^{\frac{-2D}{3\eta}(k_i^3-(k_i-\eta t)^3)}e^{-Dt_H(k_i-\eta t)^2 }\\
\times &e^{-2\gamma_k t-\gamma_k t_H}f_{in}(k_x,k_y,-t)\bar{G}.
\end{split}
\end{equation}
We consider input pulse with both Gaussian temporal and transverse profile as above, typically, $D\eta^2t_p^3, \gamma_kt_p$ are very small. Then
the total efficiency will be
\begin{equation}
\begin{split}
\varepsilon_{tot}
%&\sqrt{\frac{1}{1/\alpha + 2/\alpha ' -2}}e^{\frac{-4D\eta^2}{3}\left[\left(\frac{k_i}{\eta}\right)^3-\left(\frac{k_i}{\eta}-t_{in}\right)^3\right]}\\
%&e^{-2D\eta^2t_H
%\left(\frac{k_i}{\eta}-t_{in}\right)^2}\frac{1}{1+4D(t_H+2t_{in})/a^2}\\
=&\sqrt{\frac{1}{1/\alpha_H + 2/\alpha_{W} -2}}e^{-2\tau_{W}}e^{-2\tau_{H}}\frac{1}{1+\tau_{\perp}}\\
&+O[(D\eta^2t_p^3,\gamma_kt_p)^2]
\end{split}
\end{equation}
Typically, $\alpha_H \simeq 1, \alpha_{W} \simeq 1$, so we have
\[\varepsilon_{tot} \simeq \varepsilon_{W}\times \varepsilon_{H}\times \varepsilon_{R}\times \varepsilon_{\perp}.\]

\subsection{Efficiency optimization and estimation}
Our model did not examine other decoherence processes, such as control field-induced scattering and ground state decoherence. Our results simply quantify the effects of motional diffusion on GEM efficiency, and therefore the represent upper estimates for the performance of GEM.

\begin{figure}
\includegraphics[angle=0, width=0.5\textwidth]{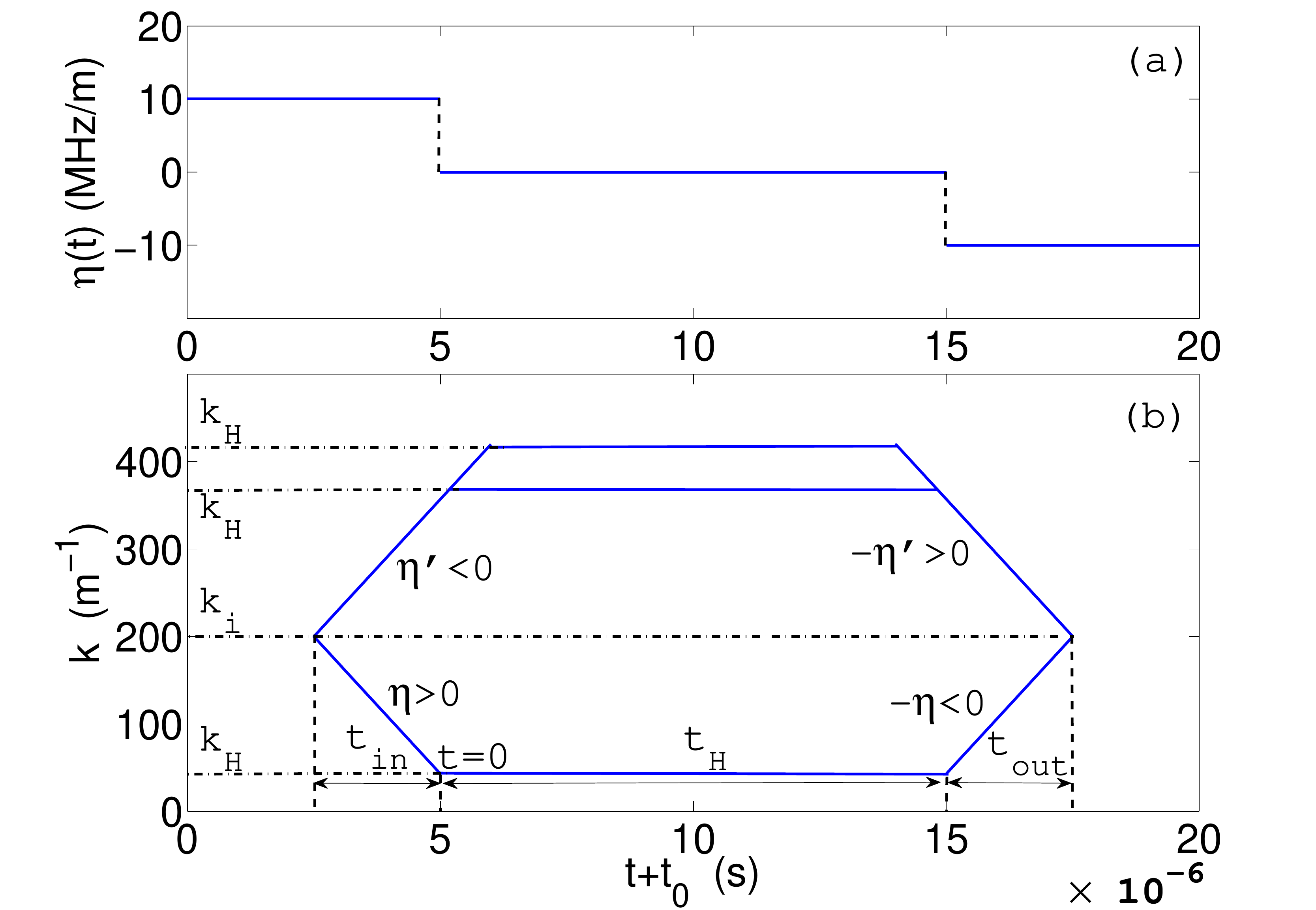}
\caption{\footnotesize{(a) The gradient is turned off during hold time, and flipped for read-out. (b) An illustration of the spatial frequency for $\sigma_{12}(z,t)$, $k$ decreases or increases depending on the sign of the gradient.}}
\label{fig:k-t}
\end{figure}

\begin{figure}
\includegraphics[angle=0, width=0.5\textwidth]{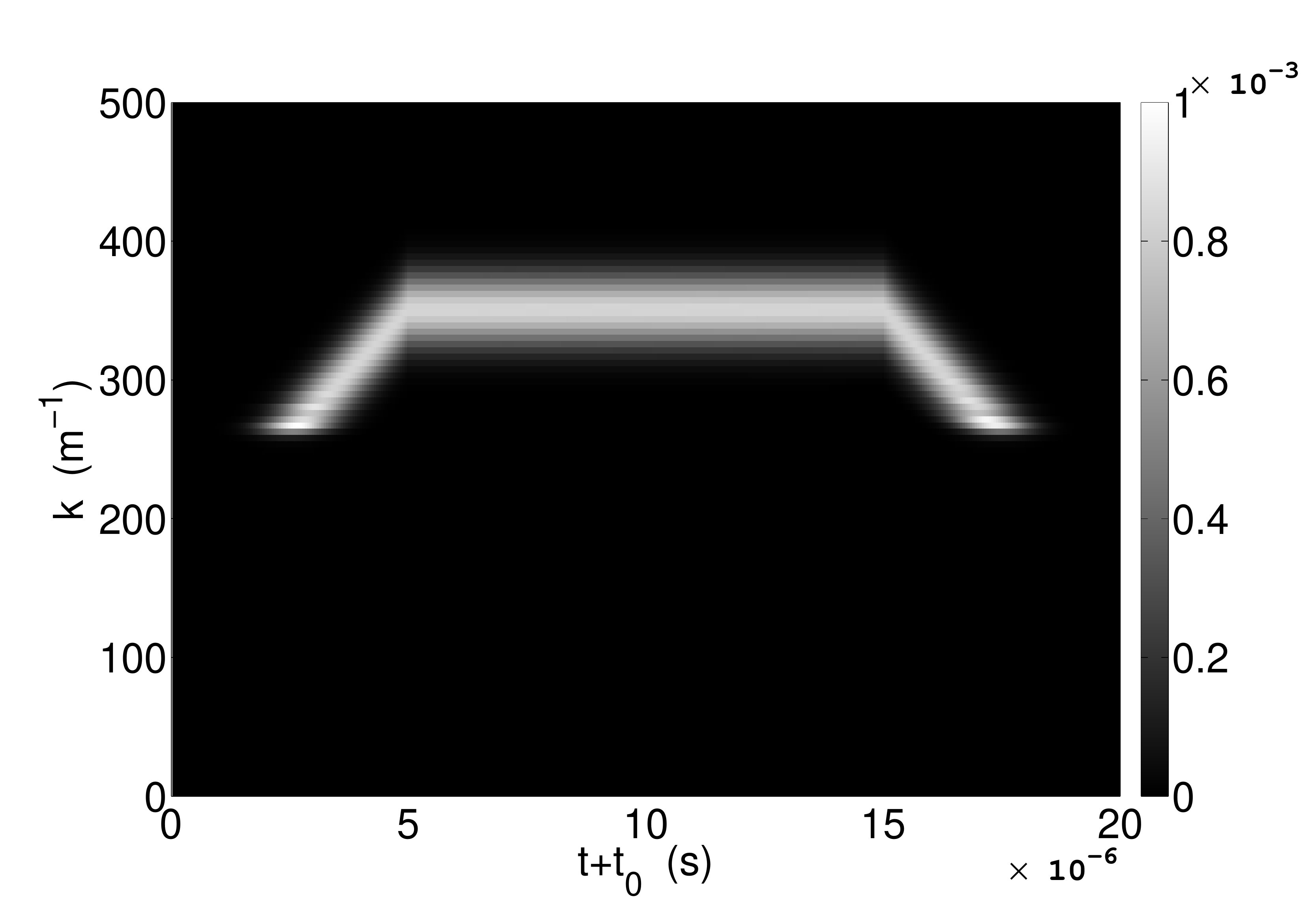}
\caption{\footnotesize{Numerical results of $|\sigma_{12}(k,t)|$ with write-in gradient $\eta'<0$.}}
\label{fig:k-t-n}
\end{figure}

Structures with larger spatial frequency will decay faster under diffusion. For the 1D model during hold time, we have (see the Appendix A)
\begin{equation}\label{eq:t=0}
\sigma_{12}(k,t)\propto f_{in}(\frac{k-k_i}{\eta})
\end{equation}
The input pulse is centered at $-t_{in}$, then we have $k\sim k_H=k_i-\eta t_{in}$. We
see that $k$ will increase or decrease as  $t_{in}$ increases, depending on the sign of $\eta$. During the hold time $[0,t_H]$, the gradient is turned off, and $k$ will hold its value. The read-out process is symmetric to the write-in process (see fig. \ref{fig:k-t}), returning the quasi-momentum to its original distribution. Fig.~\ref{fig:k-t-n} shows the numerical result by solving Maxwell-Bloch equations.

One way to reduce the effects of diffusion is to remove the gradient during the storage part of the process, and only turn on the flipped gradient during readout.
For realistic $t_{in}$ and $\eta$, we can get zero spatial frequency $k_H=0$, for which $\tau_H=0$, and this will minimise the diffusional decay rate. We then have $\varepsilon_H=\sqrt{\alpha}$ and $\varepsilon_{W}=e^{-2 D k_i^2 t_{in}/3}$. Including transverse diffusion, we get the total efficiency for input field with transverse Gaussian profile
\begin{equation}
\varepsilon_{tot} =e^{-4 D k_i^2 t_{in}/3}\sqrt{\alpha}\frac{1}{1+4D(t_H+2t_{in})/a^2}
\end{equation}
The efficiency can be improved further by choosing a larger transverse width $a$, i.e, the effects of transverse diffusion will be reduced by using a smooth field in the transverse direction.

We note that the circumstances in which a GEM will be useful are those for which all dephasing, including that due to diffusion, is small.  In this limit, a useful approximate expression for the GEM efficiency is given by
\begin{equation}
\varepsilon_{tot} \simeq 1-\frac{4Dk_i^2t_{in}}{3}-\frac{Dt_H\eta ^2 t_p^2}{2}-\frac{4D(t_H+2t_{in})}{a^2},
\end{equation}
as the inefficiencies arising from each diffusive process considered above add together.

Experimental considerations give estimates of the achievable GEM efficiency. In particular, to ensure the bandwidth of the memory is large enough to absorb the input field,  we require $|\eta t_p| > \frac{1}{L}$, and $t_{in}>t_p$ to ensure that the whole pulse enters the medium during the write-in process, also $|k_i|>\frac{1}{L}$ is required to satisfy $k_H=0$. So
\begin{equation}
\begin{split}
\varepsilon_{tot} &\lesssim 1-\frac{4Dk_i^2t_{p}}{3}-\frac{Dt_H}{2L^2}-\frac{4D(t_H+2t_{p})}{a^2}\\
&\lesssim 1-\frac{4Dt_{p}}{3L^2}-\frac{Dt_H}{2L^2}-\frac{4D(t_H+2t_{p})}{a^2}
\end{split}
\end{equation}
This gives a reasonable upper bound on the GEM efficiency, given the  pulse duration $2 t_p$, the hold time $t_H$, and the vapour length $L$ and beam width $a$.

\emph{Experimental considerations}: In  experiments reported in \cite{hosseini2011high, higginbottom2012spatial}, Rb$^{87}$ atoms were used. Typical system parameters are $\omega_0=2\pi\cdot 377.10746$ THz, $\omega_0-\omega_c=2\pi\cdot 6.8$ GHz,
$\Delta=-2\pi\cdot 1.5$ GHz, $\Omega_c\simeq 2\pi\cdot 20$ MHz, $g\simeq 2\pi\cdot 4.5$ Hz, $t_p=1 \mu$s, $a\simeq 1.45$ mm, $2L=0.2$ m, $\eta\simeq -2\pi\cdot 10$ MHz/m, $N\simeq 0.5\times 10^{18}$ m$^{-3}$ \cite{hosseini2011high, higginbottom2012spatial,steck2001rubidium}. The optical depth $|\beta|\simeq 3.8$ is sufficiently large.
According to the formula in \cite{happer1972optical}, we have $D\sim 0.004$ m$^2$/s for Rubidium atoms in buffer gas \cite{hosseini2011high, higginbottom2012spatial}.

With these parameters, the diffusive decay will be dominated by transverse diffusion.  For example, for $t_{in}=5 \mu$s and $t_H=0$, the maximum achievable efficiency is $\varepsilon_{tot}\simeq 93\%$ ($\varepsilon_{\perp}\simeq \varepsilon_{tot}$, $\varepsilon_{H}=1$,  $\varepsilon_{W}\simeq 1$).

%***Can you expand on the experimental results ***  Given the formulae are explicit, I don't think having a range of unpublished experimental parameters will be particularly interesting.

We examine the input of the (1,1) Hermite-Gaussian mode $f_{in}(x,y,t) \propto x\,y\,e^{-(x^2+y^2)/a^2}$ as an example of a higher order Hermite-Gaussian mode transverse profile.  From Eqs.~(\ref{eq:3d-eff}, \ref{eq:3d-out}), we have $\varepsilon_{\perp}=(\frac{1}{1+\tau_{\perp}})^3$, and the longitude diffusion effects are the same as the Gaussian profile (i.e. the (0,0) Hermite-Gaussian mode). Thus, for diffusive decays, we have
\begin{equation}\label{ratio}
\frac{\varepsilon_{(11)}}{\varepsilon_{(00)}} \propto (\frac{1}{1+\tau_{\perp}})^2
\end{equation}
where $\varepsilon_{(ij)}$ is the read-out efficiency for (ij) Hermite-Gaussian mode.
We find that the efficiency decays faster for higher order modes, and the ratio Eq.~(\ref{ratio}) decreases when the storage time increases.  This is in agreement with experimental investigations \cite{higginbottom2012spatial}.

\subsection{Output beam width}

After some storage time, transverse diffusion will tend to smear the spin wave density in the radial direction.  Intuitively, we would expect this to lead to a spatially wider output beam than would be the case in the absence of diffusion.

\begin{figure}
\includegraphics[angle=0, width=0.5\textwidth]{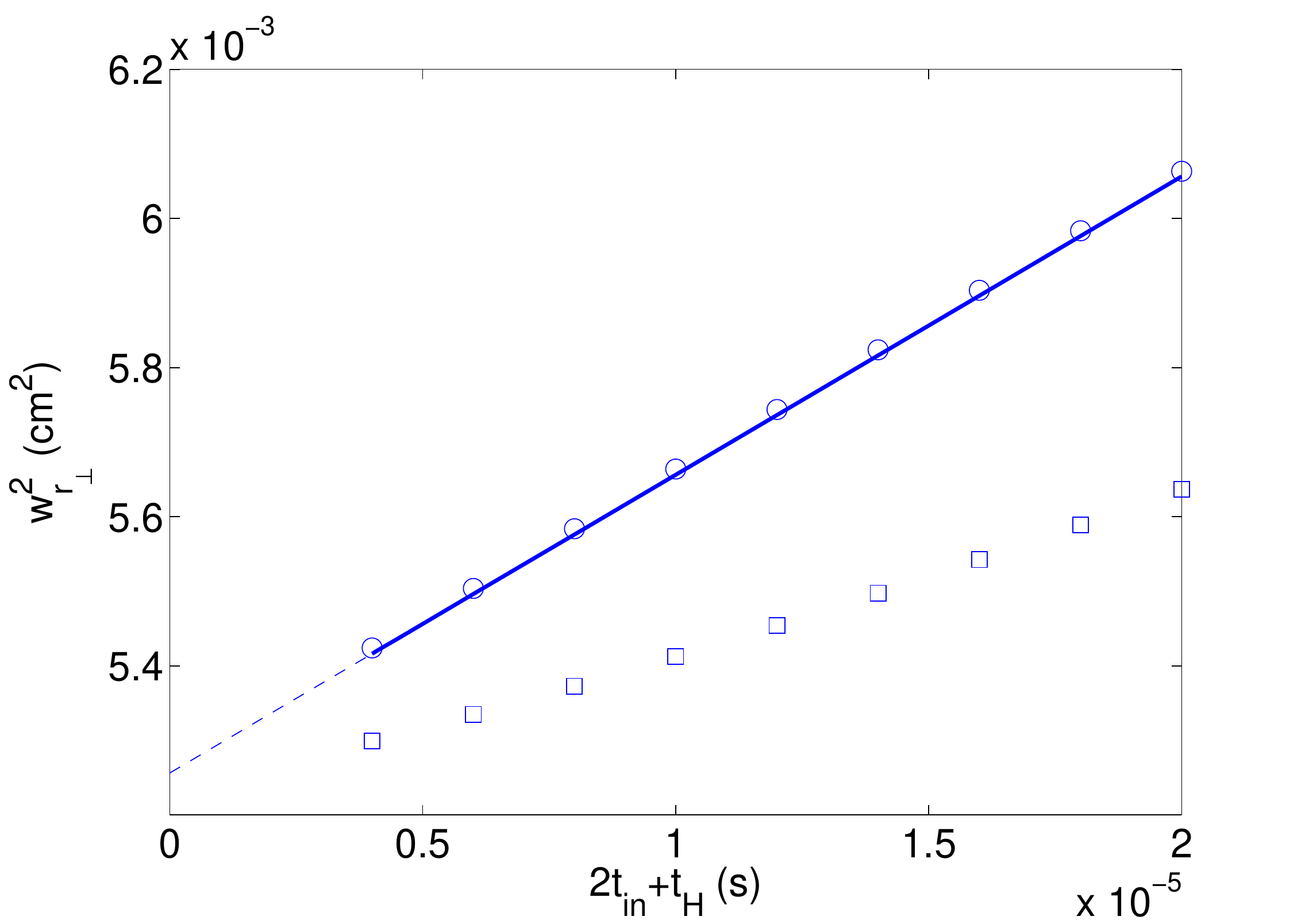}
\caption{\footnotesize{Expansion of the read-out signal. Circles are numerical results for homogeneous control field. Squares are numerical results for control field with Gaussian profile, the solid line is Eq.~(\ref{eq:width}) for homogeneous control field.}}
\label{fig:width}
\end{figure}

This is certainly the case when the control field is radially uniform.  To see this, we define the intensity distribution for the read-out signal as
\begin{equation}
I(r_{\perp})=\int |f_{out}(r_{\perp},t_H+t)|^2 dt.
\end{equation}
%Follow the experiment \cite{higginbottom2012spatial}, we consider
We suppose  that the control field is turned off during the hold time, $[0,t_H]$ to avoid control field-induced scattering, and that the gradient is always on and flipped at $t=0.5t_H$. Also for typical experimental parameters, the effects of longitudinal diffusion is very weak, so we focus on transverse diffusion. We solve the Maxwell-Bloch equation using the same method as before.  For a signal with a Gaussian transverse profile, we find that
\begin{equation}
I(r_{\perp}) \propto e^{r_{\perp}^2/[a^2+4D(2t_{in}+t_H)]}.
\end{equation}
with $r_{\perp}^2=x^2+y^2$.
Defining $w_{r_{\perp}}$ as the width of the output field, we have
\begin{equation}\label{eq:width}
w_{r_{\perp}}^2=\frac{a^2}{4}+D(2t_{in}+t_H),
\end{equation}
which increases linearly with storage time (Fig. \ref{fig:width}), at a rate determined by the diffusion coefficient.

\begin{figure}
\includegraphics[angle=0, width=0.5\textwidth]{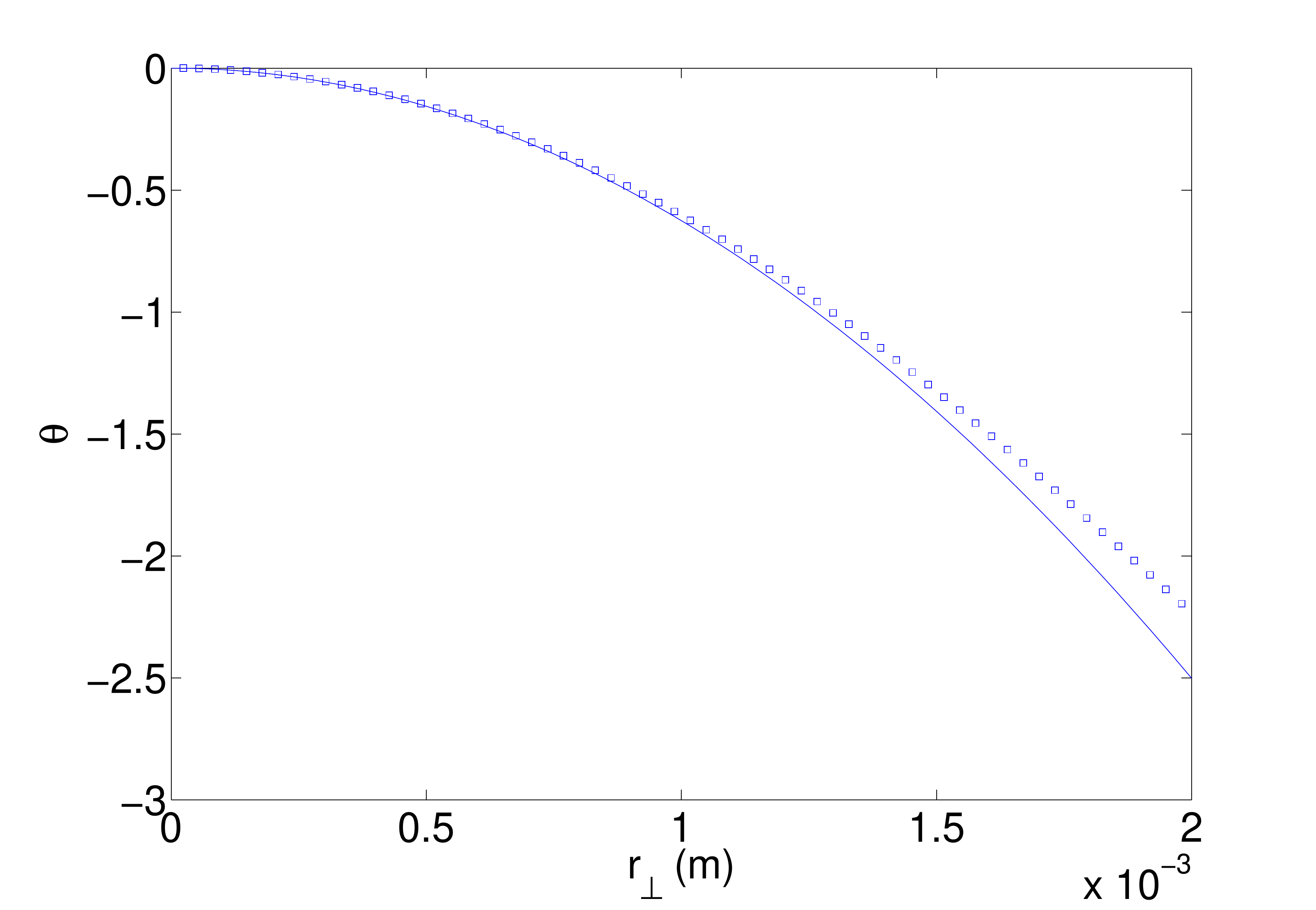}
\caption{\footnotesize{The extra phase $\theta$ for control field with Gaussian profile. Points are numerical results using typical parameters given in the main text, and the curve is the approximate expression in Eq.~(\ref{eqn:theta}).
%estimation $\theta=-0.625r_{\perp}^2$.
}}
\label{fig:phase}
\end{figure}

\begin{figure}
\includegraphics[angle=0, width=0.5\textwidth]{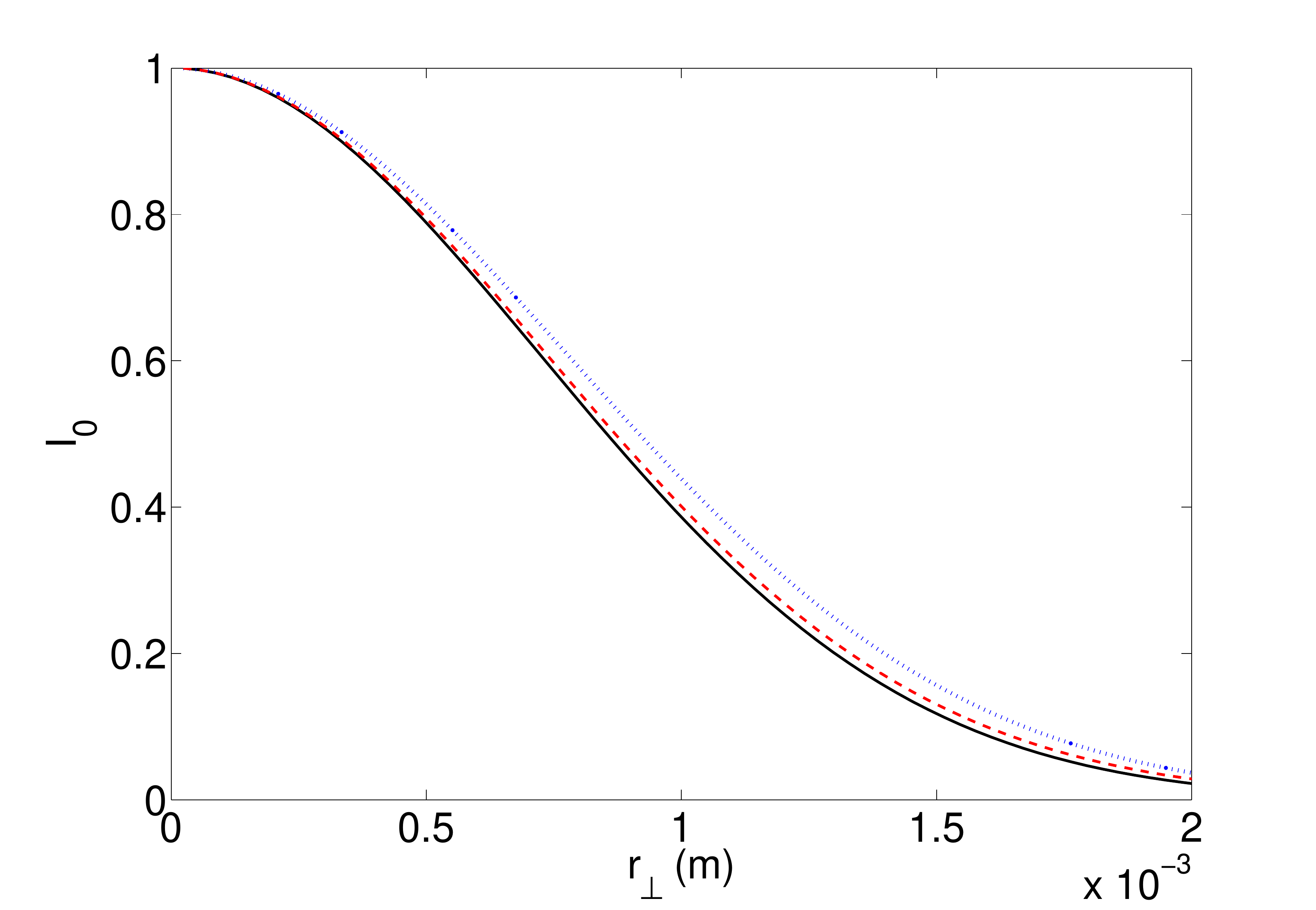}
\caption{\footnotesize{(Color online) The intensity distribution for the read-out signal with $t_H=16$ $\mu$s. To see the expansion clearly, we have renormalized the maximum of $I(r_\perp)$ to 1, and $I_0$ is the renormalized intensity distribution. The black solid curve is input signal, the blue dotted one is the read-out signal for homogeneous control field, and the red dashed is read-out signal for control field with Gaussian profile.}}
\label{fig:profile}
\end{figure}

Somewhat surprisingly, the experimentally measured rate of expansion of the read-out signal is smaller than that  expected from atomic diffusion by a factor of 2 to 3 \cite{higginbottom2012spatial}.
One possible explanation for this is the signal diffraction as suggested in \cite{higginbottom2012spatial}, diffusion leads to a beam with reduced divergence and the measurement is taken downstream. However in this experiment the scale of experimental setup is much smaller than the Rayleigh range, so the diffraction effect is  too small to explain the observed discrepancy.

Instead, we find that the anomalously narrow output beam width can be explained by considering the control field with realistic transverse Gaussian profile.  This leads to a transverse variation in the phase of the spin wave, which, under the influence of diffusion leads to lower emission efficiency in the wings of the spin wave.

To analyse this effect quantitatively, we consider a Gaussian transverse variation in the control field, $\Omega_c(r_{\perp})=\Omega e^{-r_{\perp}^2/w_c^2}$, with beam waist $w_c$. Then the two-photon detuning, $\delta$, and the optical depth becomes $r_{\perp}$ dependent.
% due to ac-Stark shift, and the optical depth $\beta$ also becomes $r_{\perp}$ dependent.

From our solution for the spin wave [see Appendix A, Eq.~(\ref{eq:t=t})], we find that the inhomogeneity of the control field intensity will introduce a transverse variation in $\delta$, which leads to a transverse dependence in the phase of the spin wave. Likewise, the transverse variation in the optical depth leads to a radially-dependent longitudinal shift in the spin wave $\sigma_{12}(\textbf{r},t)$.  In combination, these give rise to a radially dependent phase on the spin wave, $e^{i\theta(r_{\perp})}$, with the effect of the control field typically being dominant. We compare solutions for this inhomogeneous control field with solutions for the homogeneous control field to obtain the phase difference $\theta(r_{\perp})$ during hold time. Typically, the width of the control field, $w_c$, is much larger than the width of the signal field, $a$, so $\theta(r_{\perp})$ is approximately quadratic in $r_{\perp}/w_c$: % to the first order of $r_{\perp}^2/w_c^2$, we find $\theta(x,y)$ is quadratic $\theta(x,y)\propto r_{\perp}^2$.
\begin{equation}
\begin{split}
\theta(r_\perp)&=[-\frac{2\Omega ^2t_{in}}{\Delta}+2\beta \textsl{ln}\left(|\frac{\eta L t_{in}}{\beta}+1|\right)\\
&+2\beta\left(1-\frac{\Omega^2}{\Delta\eta L}\right)\left(\frac{\beta}{\eta L t_{in}+\beta}+\frac{z}{L}\right)]\frac{r_{\perp}^2}{w_c^2},
%&-\eta t_{in} \frac{\Omega ^2}{\Delta \eta}\frac{2r_{\perp} ^2}{w_c^2} \\
%&+\beta ln [|\eta L\left(t_{in}+\frac{\beta}{\eta L}\right)|]\frac{2r_{\perp} ^2}{w_c^2} \\
%&-i ln[e^{-\frac{r_{\perp} ^2}{w_c^2}}e^{-\frac{\pi|\beta|r_{\perp} ^2}{w_c^2}}\Gamma(i\beta e^{-\frac{2r_{\perp} ^2}{w_c^2}})/\Gamma(i\beta)] \\
%&+\beta^2\left(1-\frac{\Omega_c^2}{\Delta\eta L}\right)\frac{1}{\eta L\left(t_{in}+\frac{\beta}{\eta L}\right)}\frac{2r_{\perp} ^2}{w_c^2} \\
%&+\beta \frac{z}{L}\left(1-\frac{\Omega_c^2}{\Delta\eta L}\right)\frac{r_{\perp} ^2}{w_c^2}\\
\end{split}\label{eqn:theta}
\end{equation}
where $\beta$ is the optical depth corresponding to $\Omega$.
Because of this quadratic phase variation across the spin wave, diffusion acts to wash out the spin-wave coherence more quickly at larger radius, so the read-out efficiency is suppressed at larger $r_\perp$.  This will tend to reduce the apparent width of the emitted read-out signal.

\emph{Experimental considerations}: In the experimental results reported in \cite{higginbottom2012spatial}, $w_c\simeq 3$ mm, and $t_{in}\simeq 2$ $\mu$s. Using these parameters, Fig.~\ref{fig:phase} shows the transverse variation in the phase of the spin-wave at $(z=0,t=0.5t_H)$ in the absence of diffusion.  When diffusion is introduced, this transverse phase variation is smeared out, leading to reduced read-out efficiency in the wings of the spin-wave. Figure \ref{fig:profile} compares the numerical results for the expansion of the read-out signal after a specific hold time, $t_H=16~\mu$s, with a homogeneous control field (dotted, blue) and with a spatially varying control field (dashed, red), assuming the diffusion rate $D=0.004$ m$^2$/s. Figure \ref{fig:width} shows the variation in the width of the output field as a function of hold time.  We see that the expansion is slowed  for a control field with Gaussian profile (squares), compared to the case of a uniform control field (circles). Importantly, this corresponds to a reduction of the beam width expansion-rate by a factor of $~2$. The apparent diffusion rate extracted from this slower expansion rate is  $D_{eff}\simeq 0.002$ m$^2$/s.    This is quantitatively in agreement with the observations in \cite{higginbottom2012spatial}.

\section{Summary}
We have studied the effects of diffusion on the efficiency of the $\Lambda$-gradient echo memory, both numerically and analytically. We find that the efficiency is dependent on the spatial frequencies $k$ for both longitude diffusion and transverse diffusion: higher $k$ leads to more pronounced diffusive effects, and reduced efficiency, as expected.  We show that the storage efficiency can be improved by appropriate choice of the gradient during the hold phase.

We established a mechanism by which the rate of expansion of the transverse width of the beam is reduced, compared to the naive expectation of diffusive effects.  This mechanism arises from the effects of diffusion on the transverse variation in the spin wave phase.   We showed that with an experimentally reasonable choice of parameters,  the magnitude  of this effect is the same as that observed in recent experiments.  When the density of the buffer gas in increased, the collision rate increases, leading to a smaller diffusion rate.  However, this will lead to collision-induced dephasing, which will dominate at sufficiently high buffer gas pressures.  This implies a trade off between  diffusion- and collision-induced dephasing. This will be the subject of future research. %Also when other decays such as control field-induced scattering are included, the problem becomes more complicated. Further studies are necessary to address these issues.

\section*{Acknowledgments}
J. Hope and B. Hillman thank M. Hosseini, D. Higginbottom, O. Pinel and B. Buchler for helpful discussions of the modelling and experiments. J. Hope was supported by the ARC Future Fellowship Scheme.
X.-W. Luo thanks G. J. Milburn for helpful discussions, and gratefully acknowledges the National Natural Science Foundation of China (Grants No. 11174270), the National Basic Research Program of China (Grants No. 2011CB921204), CAS for financial support,  and The University of Queensland for kind hospitality.

%\bibliographystyle{unsrt}
%\bibliography{SummaryGEM}

\newpage
%\null
%\newpage
%\begin{appendices}
\appendix
\section*{Appendix A}
The Maxwell-Bloch equation for the 1-dimensional model is
\begin{equation}\label{M-B-A0}
\begin{split}
\dot{\sigma}_{12}(z,t)=&i\frac{g\Omega_c}{\Delta}e^{i(k_0-k_c)z}E(z,t)\\
&-i\delta(z,t)\sigma_{12}(z,t)+D\nabla_z^2\sigma_{12}(z,t),\\
\frac{\partial}{\partial z}E(z,t)=&
i\frac{gN\Omega_c}{c\Delta}  e^{-i(k_0-k_c)z}\sigma_{12}(z,t)\\
&+i\frac{g^2N}{c\Delta}E(z,t).
\end{split}
\end{equation}
To find the solution during $[-t_0,0]$, we first solve the equation without diffusion, then introduce the diffusion effects to our solutions.

When $D=0$, we can make transformation
\begin{equation}
\begin{split}
\widetilde{\sigma}_{12}(z,t)=&e^{-i\frac{g^2N}{c\Delta}z}e^{-i(k_0-k_c)z}\sigma_{12}(z,t),\\
\widetilde{E}(z,t)=&e^{-i\frac{g^2N}{c\Delta}z}E(z,t),
\end{split}
\end{equation}
and get the new equations
\begin{equation}\label{M-B-A}
\begin{split}
\partial_z\widetilde{E}(z,t)=&i\frac{g_{eff}N}{c}\widetilde{\sigma}_{12}(z,t),\\
\partial_t\widetilde{\sigma}_{12}(z,t)=&-i\eta z\widetilde{\sigma}_{12}(z,t)+ig_{eff}\widetilde{E}(z,t)
\end{split}
\end{equation}
where $g_{eff}=g\Omega_c/\Delta$. Following the method given in \cite{hetet2006gradient}, and using the boundary conditions $\widetilde{\sigma}_{12}(z,t\rightarrow-\infty)=0$ and $\widetilde{E}(z=-L,t<0)=\widetilde{f}_{in}(t)$, we integrate the first equation and substitute it in the second one. Making use of Fourier transformation, we find
\begin{equation}
\widetilde{E}(k,t)=\widetilde{f}_{in}(\frac{k}{\eta}+\frac{\beta}{\eta L}+t)|\frac{k}{\eta}|^{-i\beta-1}G(\eta,\beta,L),
\end{equation}
and\[G(\eta,\beta,L)=
\frac{1}{\eta}\beta e^{-\pi|\beta|/2}\textsl{sinh}(\pi|\beta|) \\
|\eta L|^{-i\beta}\Gamma(i\beta),\]
where $\widetilde{E}(k,t)=\int \widetilde{E}(z,t)e^{-ikz}dz$, $\beta=\frac{g_{eff}^2N}{\eta c}$ is the optical depth and we assume $\beta$ is sufficiently large, $\Gamma(i\beta)$ is the Gamma Function, $\widetilde{f}_{in}(t)=f_{in}(t)e^{i\frac{g^2N}{c\Delta}L}$ is the input pulse.
According to the Maxwell-Bloch equations, we have $\widetilde{\sigma}_{12}(k,t)=\frac{k\cdot c}{g_{eff}N} \widetilde{E}(k,t)$.

We transform $\widetilde{\sigma}_{12}(k,t)$ back to $\sigma_{12}(k,t)$,
\begin{equation}\label{eq:t=t}
\begin{split}
\sigma_{12}(k,t)=&f_{in}( \frac{k-k_i}{\eta}+t)e^{i\frac{g^2N}{c\Delta}L}\\
\times &|\frac{k-k_i}{\eta}-\frac{\beta}{\eta L}|^{-i\beta}\textsl{sgn}(\frac{k-k_i}{\eta}-\frac{\beta}{\eta L})\frac{c}{g_{eff}N}G
\end{split}
\end{equation}
with $k_i=\frac{g^2N}{c\Delta}+k_0-k_c-\frac{\beta}{L}$.

Now we introduce the diffusion, for the short time interval $[t,t+\Delta t]$, diffusion will cause a decay $e^{-Dk^2\Delta t}$ to $\sigma(k,t)$, or equally, there will be a decay $e^{-Dk^2\Delta t}$ on the signal $f_{in}(t')$ with $k=k_i-\eta(t-t')$. So the total decay during the write-in process for $f_{in}(t')$ is
\[
e^{-D\int_{t'}^0(k_i-\eta(t-t'))^2 dt}=e^{\frac{-D}{3\eta}(k_i^3-(k_i+\eta t')^3)}.
\]
Thus, the solution for $\sigma_{12}$ at $t=0$ is
\begin{equation}
\begin{split}
\sigma_{12}(k,0)=&e^{\frac{-D}{3\eta}(k_i^3-k^3)}f_{in}( \frac{k-k_i}{\eta})e^{i\frac{g^2N}{c\Delta}L}\\
\times &|\frac{k-k_i}{\eta}-\frac{\beta}{\eta L}|^{-i\beta}\textsl{sgn}(\frac{k-k_i}{\eta}-\frac{\beta}{\eta L})\frac{c}{g_{eff}N}G.
\end{split}
\end{equation}

We have assumed that the bandwidth of the memory is larger than the bandwidth of the input signal, $|\eta L|\gg \Delta\omega_s$, and the optical depth is sufficient large, $\beta\gtrsim 1$. The signal will be absorbed near $z=0$, and $\sigma_{12}(z,0)$ and $E(z,0)$ is nonzero only near $z=0$, so we can treat $L$ as infinity during $[0,t_H]$. Also notice that, the gradient is turned off during $[0,t_H]$, and the spatial frequency $k$ will hold its value.

To get the solution in $[0,t_H]$, we solve Eq.~(\ref{M-B-A0}) with initial condition $\sigma_{12}(k,t=0)$
for $\sigma_{12}$ and open boundary condition for $E$. In $k$ space, we find
\begin{equation}\label{f-t=t-H}
\sigma_{12}(k,t_H)=e^{-Dk^2t}e^{i\frac{g_{eff}^2N}{c}\frac{1}{k-\bar{k}}t_H}\sigma_{12}(k,0)
\end{equation}
where $\bar{k}=\frac{g^2N}{c\Delta}+k_0-k_c$.
Notice that $\sigma_{12}(k,t_H)$ get a phase $e^{i\frac{g_{eff}^2N}{c}\frac{1}{k-\bar{k}}t_H}$, so the group velocity for $\sigma_{12}(z,t)$ is $v_g(k)=\frac{g_{eff}^2N}{c(k-\bar{k})^2}$. If the memory broadening $|\eta L|$ is not much larger than the signal pulse bandwidth, the spin wave $\sigma_{12}(z,t)$ will be nonzero near the ensemble boundary. Then the spin wave will propagate to the boundary and be reflected, this may ruin the spin wave coherence near the boundary and lower the memory efficiency. One way to avoid this effect is turning off the control field during storage, which makes the effective coupling $g_{eff}=0$, and the group velocity $v_g=0$.

To find the values for $\sigma_{12}$ and $E$ in the duration $[t_H,t_H+t_0]$, one needs to solve a modified version of Eq.~(\ref{M-B-A0}) where the sign of $i\eta z$ is reversed. We follow the method given in \cite{hetet2006gradient}, propagate these equation backwards with final conditions $E(z=L,t>t_H)=f_{out}(t), \sigma_{12}(z,t\rightarrow\infty)=0$. Similar to the write-in process, at time $t_H$, we have
\begin{equation}\label{b-t=t-H}
\begin{split}
\sigma_{12}(k,t_H)=&e^{\frac{D}{3\eta}(k_i^3-k^3)}f_{out}(t_H+\frac{k-k_i}{-\eta})e^{-i\frac{g^2N}{c\Delta}L}\\
\times &|\frac{k-k_i}{\eta}-\frac{\beta}{\eta L}|^{-i\beta}\textsl{sgn}(\frac{k-k_i}{\eta}-\frac{\beta}{\eta L})\frac{c}{g_{eff}N}G^*.
\end{split}
\end{equation}
%Suppose the signal is emitted around $T+t_{out}$, then we have $k\sim k_i-\eta t_{out}$,
%so for this backwards propagating, $k$ changes in the same way as the input forwards
%process(see figs. \ref{fig:k-t}, \ref{fig:k-t-n}).
By matching the two solutions for $\sigma_{12}$ at $t_H$ Eqs. (\ref{f-t=t-H}), (\ref{b-t=t-H}), we get
\begin{equation}
\begin{split}
f_{out}(t_H+t)=&d_{W}(t)d_Hd_{R}(t)f_{in}(-t)\bar{G}
\end{split}
\end{equation}
where
\[\bar{G}=|\eta L\left(t+\frac{\beta}{\eta L}\right)|^{-i2\beta}e^{i\frac{2Lg^2N}{c\Delta}}e^{-i\frac{g_{eff}^2N}{c\eta \left(t+\frac{\beta}{\eta L}\right)}t_H}\Gamma(i\beta)/\Gamma(-i\beta)\] is a phase factor, $d_{W}(t)=e^{\frac{-D}{3\eta}(k_i^3-(k_i-\eta t)^3)}$, $d_H=e^{-D(k_i-\eta t)^2 t_H}$ and $d_{R}(t)=e^{\frac{-D}{3\eta}(k_i^3-(k_i-\eta t)^3)}$ are the diffusion decays for the write-in process $[-t_0,0]$, storage time $[0,t_H]$ and read-out process $[t_H,t_H+t_0]$ respectively.

\section*{Appendix B}
The Maxwell-Bloch equation for the 3-dimensional model is
\begin{equation}\label{M-B-tr-A}
\begin{split}
\dot{\sigma}_{12}(\textbf{r},t)=&i\frac{g\Omega_c}{\Delta}e^{i(k_0-k_c)z}E(\textbf{r},t)\\
&-(i\eta z)\sigma_{12}(\textbf{r},t)+D\nabla^2\sigma_{12}(\textbf{r},t),\\
\frac{\partial}{\partial z}E(\textbf{r},t)
=&i\frac{gN\Omega_c}{c\Delta}  e^{-i(k_0-k_c)z}\sigma_{12}(\textbf{r},t)\\
&+i\frac{g^2N}{c\Delta}E(\textbf{r},t).
\end{split}
\end{equation}
To solve these equations, we first transform transverse coordinates $x,y$ to Fourier space $k_x,k_y$,
\begin{equation}
\begin{split}
\dot{\sigma}_{12}(k_x,k_y,z,t)=&-(i\eta z+\gamma_k)\sigma_{12}(k_x,k_y,z,t)\\
&+i\frac{g\Omega_c}{\Delta}e^{i(k_0-k_c)z}E(k_x,k_y,z,t)\\
&+D\nabla_z^2\sigma_{12}(k_x,k_y,z,t),\\
\frac{\partial}{\partial z}E(k_x,k_y,z,t)
=&i\frac{gN\Omega_c}{c\Delta}  e^{-i(k_0-k_c)z}\sigma_{12}(k_x,k_y,z,t)\\
&+i\frac{g^2N}{c\Delta}E(k_x,k_y,z,t),
\end{split}
\end{equation}
where $\gamma_k=D(k_x^2+k_y^2)$. Now we make the following transformation:
\begin{equation}
\begin{split}
\bar{\sigma}_{12}(k_x,k_y,z,t)=&e^{\gamma_k t}\sigma_{12}(k_x,k_y,z,t),\\
\bar{E}(k_x,k_y,z,t)=&e^{\gamma_k t}E(k_x,k_y,z,t),
\end{split}
\end{equation}
then we have
\begin{equation}
\begin{split}
\dot{\bar{\sigma}}_{12}(k_x,k_y,z,t)=&-(i\eta z)\bar{\sigma}_{12}(k_x,k_y,z,t)\\
&+i\frac{g\Omega_c}{\Delta}e^{i(k_0-k_c)z}\bar{E}(k_x,k_y,z,t)\\
&+D\nabla_z^2\bar{\sigma}_{12}(k_x,k_y,z,t),\\
\frac{\partial}{\partial z}\bar{E}(k_x,k_y,z,t)
=&i\frac{gN\Omega_c}{c\Delta}  e^{-i(k_0-k_c)z}\bar{\sigma}_{12}(k_x,k_y,z,t)\\
&+i\frac{g^2N}{c\Delta}\bar{E}(k_x,k_y,z,t).
\end{split}
\end{equation}
These are actually quasi-1D equations, so we can
solve these equations by the method we used before, and the output field is:
\[\bar{f}_{out}(k_x,k_y,t_H+t)=d_{W}(t)d_Hd_{R}(t)\bar{f}_{in}(k_x,k_y,-t)\bar{G}.\]
We transform back to $f_{out}(k_x,k_y,t_H+t)$, and get
\begin{equation}
\begin{split}
f_{out}(k_x,k_y,t_H+t)=&d_{W}(t)d_Hd_{R}(t)\\
\times &d_{\perp}(t)f_{in}(k_x,k_y,-t)\bar{G},
\end{split}
\end{equation}
where $d_{\perp}(t)=e^{-2\gamma_k t}e^{-\gamma_k t_H}$ is the transverse difusion decay.

\end{document}